\documentclass[showpacs,amsmath,notitlepage,amssymb,aps,showkeys,floatfix,
prd,a4paper,onecolumn]{revtex4-1}

\usepackage[utf8]{inputenc} %
\usepackage{verbatim}    
\usepackage{graphicx}
\usepackage{dcolumn}
\usepackage{bm}
\usepackage{epsfig}
\usepackage{epstopdf}
\usepackage{amsfonts}
\usepackage{amssymb,amscd}
\usepackage{color} 
\usepackage{lineno} 
\usepackage[normalem]{ulem} 
\usepackage{longtable} 
\newcommand{\be}{\begin{equation}}
\newcommand{\ee}{\end{equation}}
\newcommand{\ben}{\begin{eqnarray}}
\newcommand{\een}{\end{eqnarray}}
\newcommand{\lb}{\label}

\begin{document}

\title{An effective field theory approach to monopolium}

\author{L.~M.~Abreu\footnote{luciano.abreu@ufba.br [Corresponding author]}}   

\affiliation{Instituto de F\'isica, Universidade Federal da Bahia, Campus Universit\'ario de Ondina, Salvador, Bahia, 40170-115, Brazil}

\author{M.~de~Montigny\footnote{mdemonti@ualberta.ca}}

\affiliation{Facult\'e Saint-Jean, University of Alberta, Edmonton, Alberta, T6C 4G9, Canada}

\author{P.~P.~A.~Ouimet\footnote{ouimet1p@uregina.ca}}

\affiliation{Department of Physics, University of Regina, Regina, Saskatchewan, S4S 0A2, Canada}

\begin{abstract}

In this work we investigate the interaction between spin-zero and spin-one monopoles by making use of an effective field theory based on two-body and four-body interaction parts. In particular, we analyze the formation of bound state of monopole-antimonopole (i.e. monopolium). 
The magnetic-charge conjugation symmetry is studied in analogy to the usual charge conjugation to define a particle basis, for which we find bound-state solutions  with relatively small binding energies and which allows us to identify the bounds on the parameters in the effective Lagrangians. Estimations of their masses, binding energies and scattering lengths are performed as functions of monopole masses and interaction strength in a specific renormalization scheme. We also examine the general validity of the approach and the feasibility of detecting the monopolium.

\end{abstract}

\maketitle


\section{Introduction}
\label{Introduction}

Despite the lack of experimental evidence regarding the existence of magnetic monopoles in nature, their search is reinvigorated whenever a new high energy sector becomes accessible, such as the Run 3 of the CERN LHC, expected in the Spring 2021. Through a brilliant theoretical reasoning, Dirac observed that, in addition to rendering the Maxwell equations more symmetric, the existence of a magnetic monopole would lead naturally to the discreteness of electric charge \cite{Dirac}. Whereas Dirac monopoles are allowed, but not required, to exist under a pure U(1) gauge theory, it was later observed that a non-abelian theory necessarily implies 't Hooft-Polyakov monopoles  \cite{Hooft, Polyakov}. Reviews on magnetic monopoles are in Refs.  \cite{Milton2006, Shnir}. 

Our motivation for this work is the new reach offered by the MoEDAL detector, a largely passive detector at the Interaction Point 8 on the LHC ring at CERN \cite{Moedal2014}. A recent analysis of elements of the MoEDAL trapping detector, exposed to 13 TeV LHC collisions in 2015-2017, for the presence of trapped magnetic charge taking into account the combined photon-fusion and Drell-Yan monopole-pair direct production mechanisms, provided mass limits in the range 1500-3750 GeV for magnetic charges up to 5$g_D$ for monopoles of spin 0, 1/2 and 1 \cite{Moedal2019} (see also a recent discussion from a theoretical perspective in Ref.~\cite{Baines:2018ltl}). The dimensionless quantity $g_D$ gives the magnetic charge as $q_m=ng_Dec$ where $n$ is an integer, $e$ is the electron charge, and $c$ is the speed of light in vacuum. In comparison, earlier Drell-Yan mass limits set by MoEDAL at 13 TeV ranged from 450 to 1790 GeV \cite{Moedal2018}.   In this paper, we will determine valid regions of our parameter space in relation with the monopole's mass; therefore the experimental mass limits on the monopole will help further refine the values of our parameters.

Whereas it is often suggested that an idea as elegant as  a theory of monopoles should occur in nature,  the lack of experimental evidence might suggest that monopoles will simply not manifest themselves in their trivial free form. 
Instead, they might very well be confined into a monopole-antimonopole bound state (so-called ``monopolium'') by their strong magnetic forces \cite{Dirac,Dirac2,Zeldovich,Hill,Epele, Vento,Barrie:2016wxf,Reis,Fanchiotti:2017nkk,Vento:2019auh}. Since this proposal by Dirac, the concept of monopolium has been largely explored in several contexts. For example, an application in cosmology can be seen in an early paper~\cite{Hill}, in which was discussed the general properties of a $SU(5)$ GUT monopole-antimonopole bound state. On the other hand, recent investigations of interest~ \cite{Epele,Vento,Barrie:2016wxf,Reis,Vento:2019auh} studied the feasibility of detecting monopolium in present and future colliders. In particular, following the assumption that the binding between monopole and antimonopole is much larger than that between electron and proton, Refs.~\cite{Epele,Vento,Reis,Vento:2019auh} argue that the monopole-antimonopole is so strongly bound that it has a relatively small mass, and being so potentially observable at high energy colliders and indirectly proving the existence of monopoles. 
On this subject, the authors of Ref.~\cite{Reis} computed via Breit-Wigner approach the total cross sections of ultraperipheral elastic electron-electron, proton-proton and lead-lead collisions in various colliders for the exclusive production of monopolium by photon fusion. 
Also, motivated by that conservation of magnetic charge at LHC, the authors of Ref.~\cite{Vento:2019auh} analyzed the effect of the magnetic dipole constituted of a monopole-antimonopole pair, on charged particles in collisions.  They examined how such magnetic dipoles, moving near the interaction region, would affect the particles appearing in successive bunches of the proton beam, thus providing a signal for the possible detection of monopole-antimonopole pairs, mainly in the off-forward beam direction. The analysis in Ref.~\cite{Vento:2019auh} for monopolium, which naturally depends strongly on the monopole-antimonopole potentials, suggests that the chance of detection is increased for large binding energies and small widths, and that the strong coupling limit leads to off-forward protons.
Besides, another paper that deserves attention is Ref.~\cite{Barrie:2016wxf}, where the authors made use of a modified monopole-antimonopole potential to model spin-1/2 monopole interactions, yielding two types of numerical estimations: the first one with a small monopole mass and large binding energy state that may provide a target at LHC, as in Ref.~\cite{Vento:2019auh}; and the second one with  para-$(J = 0)$ and ortho-$(J =
1)$ monopolia having degenerate masses and a small binding energy. This last situation would be observable in multi-photon emission decays at future colliders and in astrophysical observations, through an explosion or burst of $\gamma$ rays from the monopolia.

In view of the above and the fact that hitherto we have no access to the exact dynamics as well as the properties of monopoles and monopolia, we believe that there is still enough room for other contributions on this issue. Namely, we intend to offer to the community other theoretical perspective which fosters possible scenarios for monopolia, in the sense that experiments might incorporate additional features into their analyses and searches. To this end, in this work we investigate the interaction between spin-zero and spin-one monopoles by making use of an effective field theory based on two-body and four-body interaction terms. In particular, we analyze the formation of monopolium with relative small binding energies, including a discussion about eigenstates of the magnetic-charge conjugation, in analogy to the usual charge conjugation~\cite{Chang}. The inspiration and justification for the effective formalism employed here comes from the following premises: (i) if monopoles exist, it is a consensus that they would be very massive particles; (ii) since we have no accurate comprehension of their dynamics, it is reasonable to extract the basic picture of monopolia by means of a formalism in which (anti)monopole interactions are taken into account effectively through contact interactions. And looking at literature, we find a framework carrying these attributes: the so-called heavy-meson effective theory (or HMET) \cite{Wise:1992hn,Manohar,Casalbuoni:1996pg,AlFiky,Valderrama,Nieves,Abreu:2015jma,Abreu:2016dfe,Abreu:2016xlr,Liu}, defined in terms of scalar and vector meson fields, whose structure is guided by the heavy-quark, heavy flavor and chiral symmetries. In the simplified version of this model adapted to describe the monopole physics, we utilize the heavy fields representing the scalar or vector monopole and anti-monopole, with the interaction strength between them being described in terms of two parameters.
This structure allows to include naturally the magnetic-charge conjugation symmetry to define spin-one eigenstates in the particle basis for which we find bound-state solutions. We compute estimations of monopolium masses, binding energies and scattering lengths as functions of interaction strength and monopole masses.

 The paper is organized as follows. In Section \ref{Formalism}, we present the heavy-meson effective two-body and four-body Lagrangian and discuss the transition amplitudes by using the Breit approximation to relate the non-relativistic interaction potential and the scattering amplitude for spin-zero and spin-one states. We discuss the results in Section \ref{Results} and explore the conditions of existence of bound states (in the scenario of shallow bound states) in terms of parameters which determine the coupling constants and monopole's mass, allowing in principle an estimation of the monopolium's mass, binding energy and scattering lengths.  Concluding remarks are in Section \ref{Conclusions}.

\section{Formalism}
\lb{Formalism}

Because until now the properties of monopoles as well as their dynamics are not directly accessible, in this work, we employ an effective formalism alternative to the other models in order to get some insight into additional features of monopoles that could be found in particle accelerators. The plan is to explain the essential features of the interactions between spin-zero and spin-one  (anti)monopoles by four-body interactions, with the coupling constants driving the interaction strength. Besides, in view of the supposition that monopoles are very massive, it is natural to associate their dynamics to that of heavy-fields.

\subsection{Justification for the effective approach}

In the conventional Dirac approach to monopoles, the theory with which we have to deal is the Quantum Electrodynamics (QED) with strong coupling. Since the coupling constant of the interaction between monopoles $(M)$ and antimonopoles $(\overline{M})$ is large, nonperturbative and effective calculations are required. Therefore before we state the effective Lagrangian used in this work, in this subsection, we justify it as an appropriate limit of a ``fundamental theory'', i.e. we relate the dynamics engendered by contact interactions with the physics derived from Dirac scheme under certain conditions. 

We are interested in spin-zero and and spin-one (anti)monopoles. 
So, we start with a simple scalar version of QED with strong coupling, in which the electromagnetic interactions of massive spin-zero monopoles and antimonopoles are given by the magnetic dual of the standard scalar QED Lagrangian~\cite{Baines:2018ltl}, 
\ben
\mathcal{L}_{DSQED} &  = &  \left( D_\mu \phi \right )^{\dagger}
\left( D^\mu\phi \right) - m^2 \phi ^{\dagger} \phi  -
\frac{1}{4} F^{\mu\nu} F_{\mu\nu} , 
\label{lsqed}
\een
where $D_\mu \equiv \partial_\mu- i g A_\mu$, whose $g$ is the magnetic coupling of the monopole to photons, $A_{\mu}$ is the photon field, $F_{\mu\nu}\equiv \partial_\mu A_\nu -\partial_\nu A_\mu$ is the field strength tensor, and $\phi $ the monopole field. There are two interaction vertices associated with this theory: the three- and four-point vertices. Having in mind that our interest is the simplest description of the monopole-antimonopole interactions, then the $t$-channel of tree-level contribution for the $M \overline{M} \rightarrow M \overline{M}$ scattering 
is obtained keeping only the three-point vertex, yielding the following amplitude, 
\ben
i \mathcal{M} \left(  M \overline{M} \rightarrow M \overline{M} \right) &  = &  \left( -i g \right)^2 \left( p + p' \right)^{\mu} D_{\mu \nu} \left( p - p' \right) \left( q + q' \right)^{\nu}  , 
\label{ampl-dsqed}
\een
where $D_{\mu \nu} \left( k \right)$ is the photon propagator. Although it is well-known that this diagram is gauge-invariant, we choose for simplicity the Feynman gauge, giving  $D_{\mu \nu} \left( k \right) = - i \eta_{\mu \nu} / k^2$. The common potential-like approach which describes this interaction can be obtained by taking the nonrelativistic limit. This is performed using the approximations $\left( p + p' \right)\cdot \left( q + q' \right) \approx (2 m)^2$, and $\left( p - p' \right) \approx - \left( \vec{p} - \vec{p}\,' \right)$, which allows to rewrite Eq.~(\ref{ampl-dsqed}) as 
\ben
i \mathcal{M} &  \approx & - i \left( 2 m \right)^2 \frac{ g^2}{\left( \vec{p} - \vec{p}\,' \right)^2}.  
\label{ampl-dsqed-nr}
\een
Comparing the scattering amplitude in this limit to that of non-relativistic quantum mechanics (see for example Eq.~(\ref{rel1})), we have the effective potential between the monopole and antimonopole given by
\ben
V (\vec{r}) &  = & -  g^2 \int \frac{d^3 k}{(2 \pi )^2}  \frac{ e^{i \vec{k}\cdot \vec{r}}}{ |\vec{k}|^2} = -  \frac{g^2}{ 4 \pi r}. 
\label{coul-pot}
\een
We find the familiar attractive Coulomb-like potential. 

%

The potential-based approach has been employed in different analyses of the properties of monopoles. For instance, Ref.~\cite{Epele} has exploited this ``magnetic''
Coulomb potential to analyze strongly monopole-antimonopole bound states. In the present work, we compared this above-mentioned potential with another one non-singular at zero relative separation, with an exponential cut-off describing the spin-zero monopole as possessing some spatial extension, in consonance with the idea of Dirac strings~\cite{Schiff,Goebel}.
It is shown that the cut-off potential is quite close to the Coulomb potential as long as the monopole radius is larger than the classical monopole radius. Another application has been reported in Ref.~\cite{Barrie:2016wxf}, in which the monopole is assumed to have a finite-sized inner structure based on a ’t Hooft-Polyakov like solution, with the magnetic charge being uniformly distributed on the surface of a sphere. In this last formulation, the monopole and antimonopole potential becomes linear plus Coulomb outside the sphere and is constant inside. Solutions for para-and ortho-monopolia have been numerically estimated (see a discussion in Sec.~\ref{Estimation}). 

The contact interaction utilized in this work to describe the monopole-antimonopole interaction can be legitimated from the formalism above if we come back to the expression of amplitude in Eq.~(\ref{ampl-dsqed}). 
We interpret the quantity $D^{\mu \nu} (k)$ as the effective photon correlation function responsible for nonlocal term interaction, which in nonrelativistic limit gives rise to the magnetic Coulomb potential emulating the essential features of this interaction. But if we assume the following approximation for this correlation function in the coordinate space: $ g^2 D_{eff}^{\mu \nu} (x-y) \propto C \eta^{\mu \nu } \delta
(x-y)$, in momentum space, this yields $ g^2 D_{eff}^{\mu \nu} (k) \propto C \eta^{\mu \nu }$, with $C$ being a constant associated to the strength of our effective interaction. In other words, the approximation in which the effective photon correlation function gives rise to a Dirac delta simplifies a photon exchange interaction between monopoles and
antimonopoles to a contact interaction, i.e. a four-monopole point interaction. In the nonrelativistic limit, it produces a simplified constant potential. Also, this framework provides the relation between the coupling constant $C$ of the contact interaction with the Dirac strong coupling: $C \propto g^2$. 

We must notice some aspects in the formalism described in next subsection. The discussion above is similarly applicable to the case of vector monopoles. Besides, our aim is to understand the possible bound states of spin-zero and spin-one  (anti)monopoles, so we will generalize possible four-body interactions also considering coupling between them. Another relevant attribute is the supposition that monopoles are very massive, with their interactions with photons being approximately independent of the spin and their dynamics effectively associated to heavy-fields. Hence, we have the ingredients that can be put together in an effective Lagrangian for the characterization of monopolia in a perspective distinct from previous studies.

\subsection{ Effective Lagrangian}

To describe the monopole phenomenology, here we exploit a version of the HMET \cite{Wise:1992hn,Manohar,AlFiky,Valderrama,Nieves,Abreu:2015jma,Abreu:2016dfe,Abreu:2016xlr}, originally used for heavy mesons, but adapted in the present context in which the heavy fields denote the spin-zero and spin-one  monopole and anti-monopole fields.

With this in mind, at the lowest order of the heavy-field effective theory, we introduce the effective Lagrangian as
\begin{eqnarray}
\mathcal{L} = \mathcal{L}_2 + \mathcal{L}_4, 
\label{L1} 
\end{eqnarray} 
where the two-body interaction term is 
\begin{eqnarray}
\mathcal{L}_2 & = & - i \;\mathrm{Tr}\hspace{1pt} \left[ \overline{\mathcal{H}}  ^{(M) b} v 
\cdot \partial  \; \delta_{b}^{a}
 \;\mathcal{H}_{a}  ^{(M) } \right]  - i \; \mathrm{Tr}\hspace{1pt} 
 \left[ \mathcal{H}  ^{\left( \overline{M} \right)  b}
  v \cdot \partial  \; \delta_{b}^{a} \;
 \overline{\mathcal{H}} _{a} ^{\left( \overline{M} \right)  } \right]
\label{L2} 
\end{eqnarray} 
and the four-body interaction term reads
\begin{eqnarray}
\mathcal{L}_4 & = & - \frac{C_1}{4} \mathrm{Tr}\hspace{1pt}
\left[ \overline{\mathcal{H}}  ^{(M) a} 
 \mathcal{H}_{a} ^{(M) } \gamma  ^{\mu} \right] \mathrm{Tr}\hspace{1pt} \left[
 \mathcal{H}  ^{\left( \overline{M} \right) a}
 \;\overline{\mathcal{H}}_{a}  ^{\left( \overline{M} \right) } \gamma  _{\mu} \right]  \nonumber \\
& & - \frac{C_2}{4} \mathrm{Tr}\hspace{1pt}
\left[ \overline{\mathcal{H}}  ^{(M) a} 
 \mathcal{H}_{a} ^{(M) } \gamma  ^{\mu} \gamma^5 \right] \mathrm{Tr}\hspace{1pt} \left[
 \mathcal{H}  ^{\left( \overline{M} \right) a}
 \;\overline{\mathcal{H}}_{a}  ^{\left( \overline{M} \right) } \gamma  _{\mu} \gamma^5 \right]  .
  \label{L4}
  \end{eqnarray}
  
In Eqs. \eqref{L2} and \eqref{L4}, $\mathrm{Tr}$ denotes the trace over the Dirac space, with the ``superfields'' $\mathcal{H}$ given by
\begin{eqnarray}
\mathcal{H}_a ^{(M)} & = & \left( \frac{1+ v_{\mu} \gamma ^{\mu} }{2}\right)\left( 
M _{a \mu}^{*} 
\gamma ^{\mu}  - M _{a }  \gamma ^{5} \right) ,\nonumber \\
\mathcal{H} ^{ \left( \overline{M} \right)  a} & = & \left( \overline{M} _{ \mu}^{*  a} \gamma ^{\mu} - 
\overline{M} ^{a}  \gamma^{5} \right) \left( \frac{1 - v_{\mu} 
\gamma ^{\mu}}{2}\right). 
\label{H2}
\end{eqnarray}
In Eq. (\ref{H2}), $v$ is the velocity parameter, $\gamma$ are the usual Dirac matrices, $a$ is the index of  a flavour-like group  $SU(N)_V$ corresponding to possible different types of monopoles. The fields $M _{a } \left( \overline{M} ^{a } \right)$ and $M _{a \mu}^{* } \left( \overline{M} _{ \mu}^{* a} \right) $ are, respectively, the scalar and vector heavy fields associated to the monopoles (anti-monopoles) [we use different notation from previous subsection].
Notice that the vector fields obey the conditions: 
\begin{eqnarray}
 v \cdot  M _{a }^{* } & = & 0, \nonumber \\
  v \cdot  \overline{M} _{ \mu}^{*  a} & = & 0,
\label{C1}  
 \end{eqnarray}
which define the three different polarizations of the heavy vector fields.
In order to construct invariant quantities under the symmetry discussed above, we also used in Eqs. (\ref{L2}) and (\ref{L4}) the hermitian conjugate fields,
\begin{eqnarray}
\overline{\mathcal{H}}  ^{(M) a} & = & \gamma ^0 \mathcal{H}_a  ^{(M) \dagger } \gamma ^0 ,
 \nonumber \\
\overline{\mathcal{H}}_a  ^{\left( \overline{M} \right) }  & = & \gamma ^0 \mathcal{H}  ^{(Q) a \dagger }
 \gamma ^0.
\label{H7}
\end{eqnarray}
Note that under $SU(N)_V$-flavor representation $\mathcal{U}_{ a b }$, the superfields transform as
  \begin{eqnarray}
     \mathcal{H}_a  ^{(M)} & \rightarrow &  \mathcal{U}_{ a b }
  \mathcal{H}_b  ^{\left( M \right) } , \nonumber \\
  \mathcal{H} ^{(\overline{M}) a} & \rightarrow &    
  \mathcal{H}  ^{ \left(\overline{M} \right)   b} \mathcal{U}^{\dagger b a } , \nonumber \\
  \overline{\mathcal{H}}  ^{(M) a} & \rightarrow & \overline{\mathcal{H}}  ^{\left( M \right)  b} \mathcal{U}^{\dagger b a } , \nonumber \\
  \overline{\mathcal{H}}_a ^{\left(\overline{M} \right) } & \rightarrow &    
  \mathcal{U}_{ a b }  \overline{\mathcal{H}} _b ^{\left(\overline{M} \right) }.\label{H8}
  \end{eqnarray}
  
It is worth mentioning that, in principle, it is possible to construct other Lorentz-invariant structures at the leading order.  However, as remarked in Ref. \cite{Liu}, these other contact terms are not  independent, but are linear combinations of terms in Eq. (\ref{L4}). Thus, we will omit them.  Besides, concerning the $1/m_M$ expansion, where $m_M $ is the monopole's mass, we work at the leading order; relativistic effects are suppressed and a system constituted of two heavy particles can be described in the non-relativistic version of the theory. Therefore, it is more convenient to adopt the rest frame of the heavy particle, i.e. the velocity parameter is chosen as $v = \left( 1, \vec{0} \right)$. Also, the following normalization is employed
\cite{Manohar,Valderrama,Abreu:2015jma}, 
\be 
  \sqrt{2}  M_{a }^{(* \mu)} \rightarrow   M_{a }^{(* \mu) }. 
\label{norm}
\ee
Observe that the notation above renders the component $\mu =0$ of the vector field irrelevant, so that, henceforth, it will be sufficient to deal only with the Euclidian part of the vector heavy fields. 

Next we perform the expansion of the $\mathcal{H}$-fields in the heavy-particle limit, so that the four-body interaction terms in Eq. (\ref{L4}) read 
\begin{eqnarray}
\mathcal{L}_4 & = & C_1 \left(  M^{*  a \dagger } \cdot  M _{a}^{ * }   
+ M^{  a \dagger }   M _{a} \right)  
\left(   \overline{M}^{*  a^{\prime} } \cdot  \overline{M}_{a^{\prime}}^{ * \dagger}   
+ \overline{M}^{ a^{\prime} }   \overline{M}_{a^{\prime}}^{\dagger}  \right)
\nonumber \\
& & - C_2 \left( M^{*  a \dagger } \times  M _{a}^{ * }  \right) 
\cdot \left(  \overline{M}^{*  a  } \times \overline{M}_{a}^{ * \dagger} \right)  \nonumber \\
& & - i C_2 \left[ \left( M^{*  a \dagger } \times  M _{a}^{ * }    \right) 
\cdot  \left( \overline{M}^{*  a^{\prime} }  \overline{M}_{a^{\prime}}^{ \dagger}  + \overline{M}^{ a^{\prime} }   \overline{M}_{a^{\prime}}^{* \dagger} \right)
\right. \nonumber \\
& & + \left. \left( M^{*  a \dagger } M _{a}   
+ M^{  a \dagger }   M _{a}^{*} \right) \cdot 
\left(  \overline{M}^{*  a  } \times \overline{M}_{a}^{ * \dagger}  \right) \right] \nonumber \\
& &  - C_2 \left(  M^{*  a \dagger }   M _{a}   
+ M^{  a \dagger }   M _{a}^{*} \right) \cdot
 \left( \overline{M}^{*  a^{\prime} }  \overline{M}_{a^{\prime}}^{ \dagger}  + \overline{M}^{ a^{\prime} }   \overline{M}_{a^{\prime}}^{* \dagger} \right).
 \label{L4mod}
 \end{eqnarray}
The polarization of the vector fields $M_a ^{ *}$ and $\overline{M}_a ^{ *}$ and the sum over the $a , a^{\prime} $ indices are implicitly considered. Therefore, the interaction strength is described by  two parameters: $C_1 $ and $ C_2$.  With these definitions, in the  next section, we will discuss the existence of bound states in the present context.

\subsection{Transition Amplitudes}

Let us discuss the scattering
\be 
 M^{(*)}(1)   \overline{M}^{(*)} (2) \rightarrow M^{(*)}(3)  \overline{M}^{(*)} (4). 
\label{scat}
\ee
We use the Breit approximation to relate the non-relativistic 
interaction potential, $V$, and the scattering amplitude $ i \mathcal{M}
\left( M^{(*)} \overline{M}^{(*)} \rightarrow M^{(*)} \overline{M}^{(*)} \right) $,
\ben 
V(\vec{p}) = - \frac{1}{\sqrt{\Pi _i 2 m_i \Pi _f 2 m_f}} \mathcal{M} \left( M^{(*)}   \overline{M}^{(*)} \rightarrow M^{(*)} \overline{M}^{(*)} \right),
\label{rel1} 
\een
where $m_i $ and $m_f$ are the masses of initial and final states, and $\vec{p}$ is the momentum exchanged between the particles in the center-of-mass frame. 

From the four-body interaction term $\mathcal{L}_4$ in Eq. (\ref{L4mod}), we can determine the scattering amplitude at the tree-level approximation in terms of the states $  | M  \overline{M} \rangle, \, | M^{*}  \overline{M} \rangle, \, | M  \overline{M}^{*} \rangle, \, | M^{*}  \overline{M}^{*} \rangle  $. The resulting effective potential $V$ is shown in Table 
\ref{table1}. Although our approach permits one to treat $N$ types of monopoles, for simplicity, we shall restrict ourselves to the case $N=1$. 

\begin{table}[ht]
  \caption{Terms of interaction potential $V (\vec{p})$ in the basis 
$\left\{ |  M  \overline{M} \rangle, \, | M^{*}  \overline{M} \rangle, \, | M  \overline{M}^{*} \rangle, \, | M^{*}  \overline{M}^{*} \rangle \right\}  $; $\vec{\varepsilon}_i$ denotes the polarization 
of incoming or outgoing vector field; $\vec{S}_i$ is the spin-one operator, 
whose matrix elements are equivalent to the vector product of polarizations: 
$\vec{S}_1 \equiv  (\vec{\varepsilon}_3 ^{*}  \times \vec{\varepsilon}_1); \;\;  
\vec{S}_2 \equiv  (\vec{\varepsilon}_4 ^{*}  \times \vec{\varepsilon}_2).
$ }
\begin{center}
\begin{tabular}{|c|c|c|c|c|} \hline \hline
 & $  M  \overline{M} $ & $ M^{*}  \overline{M} $ &  $ M  \overline{M}^{*} $  &  $ M^{*}  \overline{M}^{*} $ \\  \hline \hline
$  M  \overline{M} $ & $C_1$  &  0    &    0   & 
$- C_2 \vec{\varepsilon}_1 \cdot \vec{\varepsilon}_2 $\\
$ M^{*}  \overline{M} $ & 0  & $C_1 \vec{\varepsilon}_3 ^{*} \cdot \vec{\varepsilon}_1$ 
   & $- C_2 \vec{\varepsilon}_3 ^{*} \cdot \vec{\varepsilon}_2$
& $- C_2 \vec{\varepsilon}_2 \cdot \vec{S}_1$ \\
$ M  \overline{M}^{*} $ & 0  &  $- C_2 \vec{\varepsilon}_4 ^{*} \cdot \vec{\varepsilon}_1 $  
 & $ C_1 \vec{\varepsilon}_4 ^{*} \cdot \vec{\varepsilon}_2 $
 & $ C_2 \vec{\varepsilon}_1 \cdot \vec{S}_2$ \\
 $ M^{*}  \overline{M}^{*} $ & $- C_2 \vec{\varepsilon}_3  ^{*}\cdot \vec{\varepsilon}_4 ^{*}$ 
&  $- C_2 \vec{\varepsilon}_4 ^{*} \cdot \vec{S}_1$
& $C_2 \vec{\varepsilon}_3 ^{*} \cdot \vec{S}_2$ 
& $C_1 \vec{\varepsilon}_3 ^{*} \cdot \vec{\varepsilon}_1
\vec{\varepsilon}_4 ^{*} \cdot \vec{\varepsilon}_2 + C_2\vec{S}_1 \cdot 
\vec{S}_2$ \\
 \hline
\end{tabular}
\end{center}
\label{table1}
\end{table}

The dynamically generated poles are obtained through the transition amplitudes via the Lippmann-Schwinger equation,
\ben
T ^{(\alpha \beta)} = V ^{(\alpha \beta)} + \int \frac{d^{4}q}{(2 \pi )^{4}} 
V^{(\alpha \gamma)} \,G^{(\gamma \delta)} \, T ^{(\delta \beta)}, 
\label{LS1}
\een
where $\alpha, \beta, \gamma, \delta \equiv | M  \overline{M} \rangle, \, | M^{*}  \overline{M} \rangle, \, | M  \overline{M}^{*} \rangle, \, | M^{*}  \overline{M}^{*} \rangle $,  and $G$ is given by 
\ben 
G \equiv
\frac{1}{\frac{\vec{p}^{2}}{2 m_{M^{(*)}} } + 
q_0 - \frac{\vec{q}^{2}}{2 m_{M^{(*)}} } + i \epsilon} \;\;
\frac{1}{\frac{\vec{p}^{2}}{2 m_{M^{(*)}} } + 
q_0 - \frac{\vec{q}^{2}}{2 m_{M^{(*)}}}  + i \epsilon }. 
\label{G1}
\een
Then,  we can write solutions of the Lippmann-Schwinger equation for a specific channel as 
\ben 
T ^{ (\alpha) } = \frac{V ^{ (\alpha) } }{1 - V ^{ (\alpha) } G ^{ (\alpha) } }. 
\label{LS2}
\een

For our purpose, it is sufficient to identify the pole structure of the Lippmann-Schwinger equation as follows: resonances are associated to the poles located in the second Riemann sheet (i.e. the fourth quadrant of the momentum complex plane), whereas bound states are in the first Riemann sheet (below the threshold). Since we are interested in bound-state solutions (monopolium-type solutions), from the residue theorem and dimensional regularization, Eq. (\ref{LS2}) becomes \cite{AlFiky} 
\ben 
T ^{ (\alpha) } = \frac{\tilde{V} ^{ (\alpha) } }{1 + \frac{i}{8 \pi} 
\mu |\vec{p}| \;\tilde{V} ^{ (\alpha) } },
\label{CS3}
\een
where $\tilde{V} ^{ (\alpha) }$ is the renormalized potential (or  renormalized contact interaction), and  $\mu $ is the reduced mass of the $M^{(*)} \overline{M}^{(*)}$ system, which is dependent of the  renormalization scheme. 

From Eq. (\ref{CS3}), we can compute relevant physical quantities such as the binding energy, 
\ben 
E_b ^{ (\alpha) } = \frac{32 \pi^{2} }{\left( \tilde{V} ^{ (\alpha) } \right)^2 
\mu ^3 }, 
\label{BE1}
\een
and the scattering length, 
\ben 
a_s ^{ (\alpha) } = \frac{ \mu \tilde{V} ^{ (\alpha)  }  }
{ 8 \pi } .
\label{SL}
\een
 It is worth mentioning that despite the renormalization dependence of  $\tilde{V} ^{ (\alpha)}$, quantities such as binding energy and scattering length are in principle observables, and therefore renormalization-independent. 

\subsection{Particle Basis}
\lb{Basis}

It is convenient to rewrite the four states used to describe the terms of interaction potential shown in Table~\ref{table1} according to a particle basis, that is, 
$\mathcal{B} \equiv \left\{ |  M  \overline{M} \rangle, \, | X_{+} \rangle, | X_{-} \rangle, \, | M^{*}  \overline{M}^{*} \rangle \right\}  $, where the spin-one states $X_{\pm}$ are defined as 
\be 
| X_{\pm} \rangle \equiv \frac{1}{\sqrt{2}} \left[ | M^{*}  \overline{M} \rangle \pm | M  \overline{M}^{*} \rangle \right]. 
\lb{Xpm}
\ee
In this context, $| X_{\pm} \rangle $ appear as eigenstates of magnetic--charge $(G_M)$ conjugation~\cite{Chang}. In other words: similarly to the usual operation of charge conjugation, they yield positive or negative eigenvalues with respect to the conjugation of magnetic charge of the monopole. Hence, the quantum numbers of the states defined in the $\mathcal{B}$-basis are summarized as $J^{ G_M} = 0 ^{ + }$, $1^{ \pm }$, $2 ^{ + }.$

Based on the potential terms condensed in Table~\ref{table1} and Eqs.~(\ref{BE1}) and~(\ref{SL}), we now analyze the specific channels which admit bound state solutions. 

\subsubsection{$ M  \overline{M} $ system\lb{2c1}}

This state consists of two spin-zero monopoles and is denoted as $({}^1 S_ 0 )$ in spectroscopic notation, whose bound state solution, from Table~\ref{table1} and  Eqs.~(\ref{BE1}) and~(\ref{SL}), is obtained with 
\be 
\tilde{V} ^{ ( M  \overline{M}) } = \tilde{C}_1. 
\lb{spin0}
\ee
In this specific channel, $m_M = m_{\overline{M}}$, so that the binding energy is
\ben 
E_b ^{ (M  \overline{M}) } = \frac{256 \pi^{2} }{ \tilde{C}_1 ^2 m_M ^3 }, 
\label{BE1-spin0}
\een
and the scattering length, 
\ben 
a_s ^{ (\alpha) } = \frac{ m_M \tilde{C}_1  }
{ 16 \pi } .
\label{SL-spin0}
\een
Hence, the state ${}^1 S_ 0$ depends only on one parameter, $\tilde{C}_1 $.  

\subsubsection{$ M^{*}  \overline{M}^{*}  $ system}

The combination of two spin-one monopoles leads to the states ${}^1 S_ 0$, ${}^3 S_ 1$ and ${}^5 S_ 2$. Therefore, proceeding as in Section \ref{2c1}, the bound state solutions lead to
\ben 
\tilde{V} ^{ ( M^{*}  \overline{M}^{*} [{}^1 S_ 0] ) } & = & - \tilde{C}_1 + 2 \tilde{C}_2, \nonumber \\ 
\tilde{V} ^{ ( M^{*}  \overline{M}^{*} [{}^3 S_ 1] ) } & = &  - \tilde{C}_1 +  \tilde{C}_2, \nonumber \\ 
\tilde{V} ^{ ( M^{*}  \overline{M}^{*} [{}^5 S_ 2] ) } & = &  - \tilde{C}_1 - \tilde{C}_2.
\lb{spin2}
\een
We obtain expressions similar to Eqs.~(\ref{BE1-spin0}) and (\ref{SL-spin0}).

\subsubsection{$ X_{\pm} \,\, [{}^1 S_ 0]$ systems}

From Table~\ref{table1}, one can see that we have a system of Lipmann-Schwinger equations, due to different possibilities of combinations of loop contributions.  First, we seek the bound-state solution for the state $ X_{+} $. 
If we define $T_{i j } \equiv \langle i | T | j \rangle $, where $i,j = M^{*} \overline{M} , M \overline{M}^{*}$, we obtain a system of four Lipmann-Schwinger equations which, after diagonalization, has a solution of the form
\be
T_{++}  \equiv   \langle X_{+} | T |  X_{+} \rangle 
 =   \frac{1}{2} \left[ T_{11} + T_{12}  + T_{21} + T_{22} \right] 
 =  \frac{V ^{ ( X_{+}) } }{1 - V ^{ ( X_{+}) } G ^{ (M^{*}  \overline{M}) } },
\label{LS2-XPlus}
\ee
where  
\ben 
\tilde{V} ^{ ( X_{+} ) } = - \tilde{C}_1 + \tilde{C}_2.
\lb{Xplus}
\een
The binding energy and scattering length have expressions similar to Eqs.~(\ref{BE1}) and (\ref{SL}). 

On the other hand, the orthogonal solution $T_{- -}$ is 
\be
T_{- -}  \equiv   \langle X_{-} | T |  X_{-} \rangle
 =   \frac{1}{2} \left[ T_{11} - T_{12}  + T_{21} - T_{22} \right] 
 =  \frac{V ^{ ( X_{-}) } }{1 - V ^{ ( X_{-}) } G ^{ (M^{*}  \overline{M}) } },
\label{LS2-XMinus}
\ee
where  
\ben 
\tilde{V} ^{ ( X_{-} ) } = - \tilde{C}_1 - \tilde{C}_2.
\lb{Xminus}
\een

In next section, we study the monopolium state solutions by analyzing the $(\tilde{C}_1 , \tilde{C}_2)$ space parameter and their dependence on the monopole's mass.  
In order to simplify the notation, hereafter, we shall denote the renormalized parameters simply by $C_1$ and $ C_2$.

\section{Results}
\lb{Results}

In this section, we analyze the possible bound states of $M^{(*)} \overline{M}^{(*)}$ systems discussed above. First, it is important to spend attention to the validity of our approach. This analysis must be restricted to the region of relevance of contact-range interaction, that is, the region where contributions coming from one-particle exchange is not relevant. In this sense, we study bound states which, in principle, obey the following requirement for the scattering length: $a_s \gg \lambda _{P}$, where $\lambda _{P} = 1/ m_{P}$ is the Compton wavelength of possible exchanged particles. Notwithstanding, we emphasize that the precise monopole dynamics remains unaccessible, so we should see this condition as a general validity constraint for  the model but without an exact limit. 

Besides, the lack of experimental evidence of monopoles and monopolia engenders a natural issue about the bounds on the monopole's masses $m_M$ and $m_{M^{*}}$ to estimate the binding energy and scattering length. But in light of calculations of previous Section,  $E_b$ and $a_s$ depend on couplings and reduced mass in a distinct way: the increase of both $\tilde{V} ^{ (\alpha) }$ and $\mu$ drops $E_b$ but augments $a_s$. Otherwise stated, greater values of scattering length mean  smaller values of binding energy. For that reason, it seems for us reasonable to pick out some range of validity for the binding energy according to the bound in $a_s $ as specified by last paragraph. Therefore, we determine the $(C_1 , C_2)$-space parameter regions and their dependence with the monopoles' masses which admit monopolium solutions in the scenario of relatively small bound states, i.e. by fixing the binding energies in the range $0.01 m_{M^{(*)}} \leq E_b \leq 0.1 m_{M^{(*)}}$.

Note also that we work in a specific renormalization scheme, in which the relation between the bare coupling constants is conserved after renormalization, despite the scheme dependence of each sector. Therefore, the different sectors can be related according to their dependence with the renormalized coupling constants.  Thus, Eqs. (\ref{BE1-spin0})-(\ref{LS2-XMinus}) allow us to estimate the binding energy and scattering length of $S$-wave bound states as functions of interaction strengths.

\subsection{$ M  \overline{M} ({}^1 S_0)$ system}

We begin by analyzing the $ M  \overline{M} ({}^1 S_0)$ system state. As shown in Eq.~(\ref{BE1-spin0}), the monopolium solution depends only on the parameter  
$C_1$, so  that we have to explore the $(C_1, m_M)$-space. 

\begin{figure}[htbp] 
	\centering
	\includegraphics[width=0.5\textwidth]{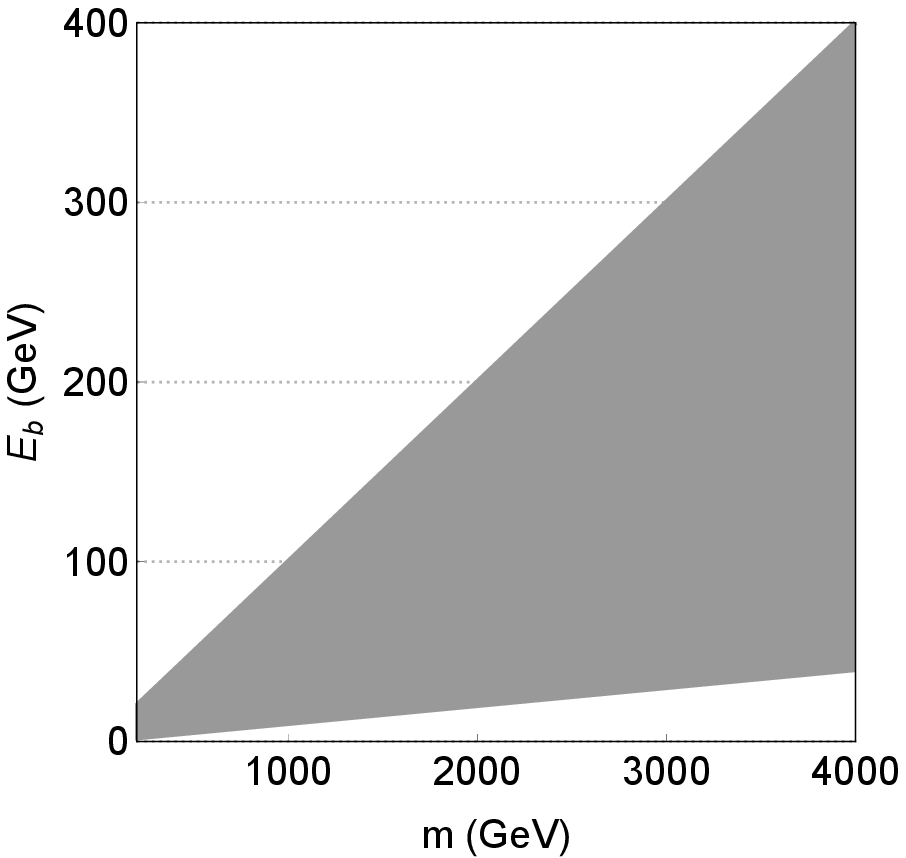}\\
	\includegraphics[width=0.5\textwidth]{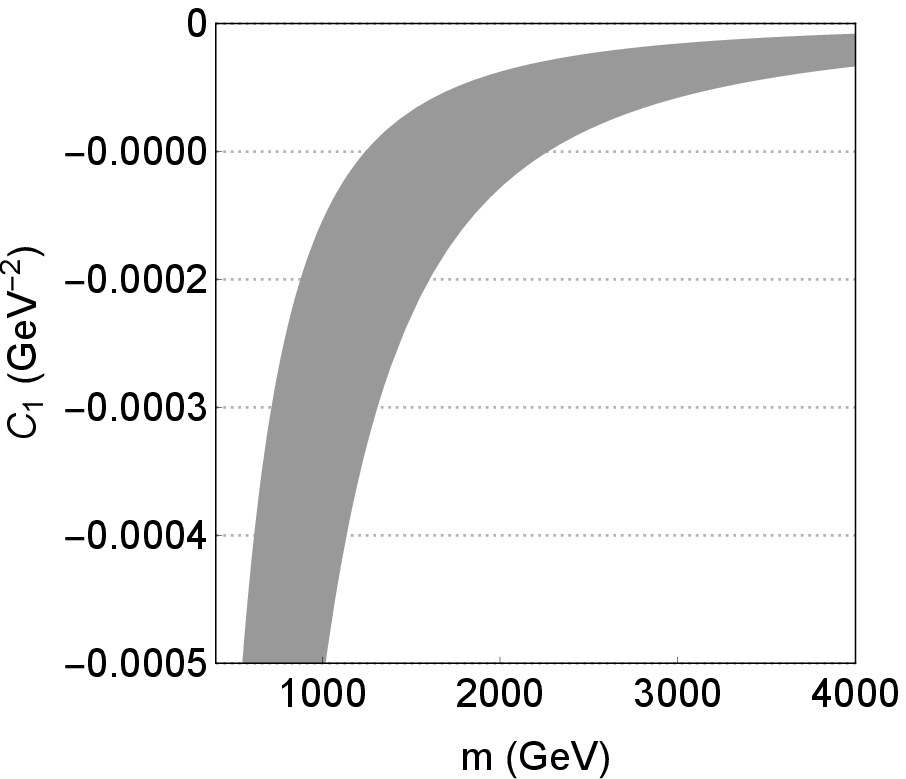}\\	\includegraphics[width=0.5\textwidth]{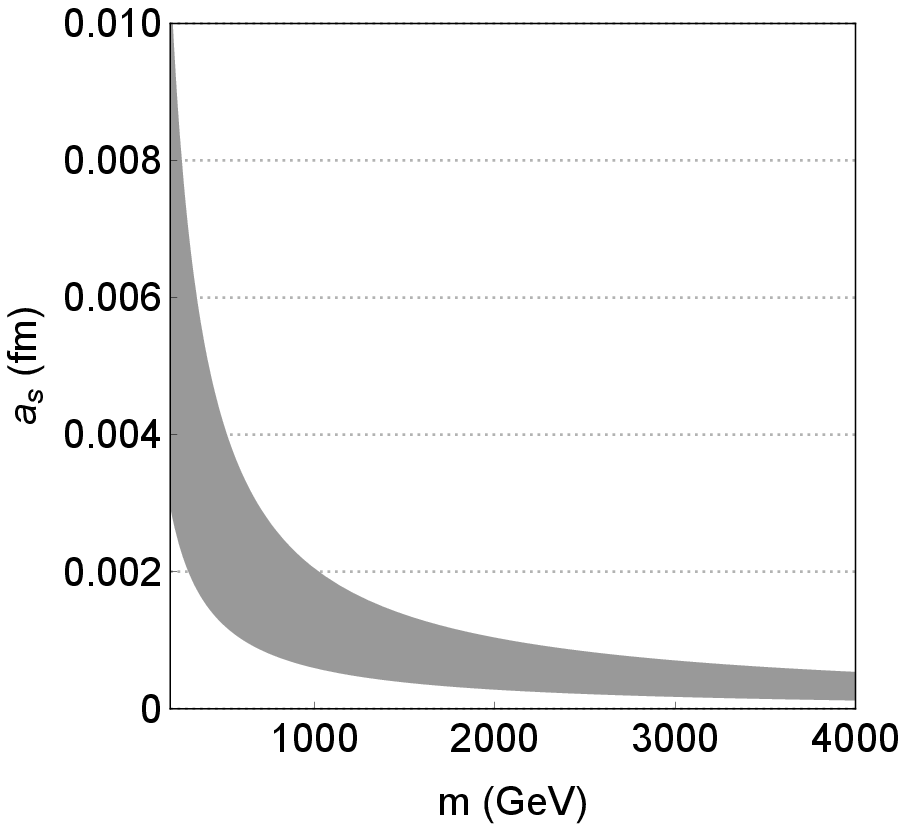}
\caption{Top panel: dependence of monopolium binding energy with the monopole mass $m_M$ for the $ M  \overline{M} ({}^1 S_0)$ system; shaded area represents the region with $ 0.01 m_{M^{(*)}} \leq E_b\leq 0.1 m_{M^{(*)}}$ (i.e. shallow bound state solutions).  Middle panel: $(C_1, m_M)$-space; shaded area represents the region in which the parameter $C_1$ acquires values that allow shallow bound state for the $ M  \overline{M} ({}^1 S_0)$ system. Bottom panel: dependence of monopolium scattering length with the monopole mass $m_M$ for the $ M  \overline{M} ({}^1 S_0)$ system;   shaded area represents the region with shallow bound state solutions. }
	\label{FIG-spin0}
\end{figure}

In Fig. \ref{FIG-spin0}, we plot the dependence of the monopolium's binding energy and scattering length on the monopole's mass $m_M$, for the $ M  \overline{M} ({}^1 S_0)$ system. Shaded areas represent the region with $0.01 m_{M}\leq E_b\leq 0.1 m_{M}$,  i.e. region in which shallow bound state solutions are obtained. Also, for completeness, we show specifically the region in which the parameter $C_1$ acquire values that allow a shallow bound state for the $ M  \overline{M} ({}^1 S_0)$ system. 

We see that monopolium solutions for the $ M  \overline{M} ({}^1 S_0)$ system are admitted only for $C_1 < 0$. Also, if we increase the monopole's mass, the scattering length diminishes. For greater binding energies, we find smaller scattering lengths. This analysis is an important feature since the validity of the contact interaction used in this present approach requires relatively large values of $a_s$. Just for the sake of comparison, in usual HMET, the region of relevance of contact-range interaction is that where the pion-exchange contribution is not relevant. In this sense, bound states must obey the requirement $a_s \gg \lambda _{\pi}$, where $\lambda _{\pi } = 1/ m_{\pi}
\sim 1 $ fm is the pion's Compton wavelength. Thus, our findings suggest a much smaller scattering length in the context of monopoles, yielding the validity domain of contact interaction in situations which exchanged particles are much heavier than the pion.  Note that this approach is valid only in the context of large values of $a_s$, when compared to Compton wavelength of exchanged particle $(\lambda_{ex} = 1 / m_{ex})$. Therefore, it seems that for a very massive monopole, $a_s$ would be very small, but if the exchanged particle is also very massive, the condition $a_s > \lambda_{ex} $ is satisfied.

\subsection{$M^{*}  \overline{M}^{*} ({}^1 S_0), M^{*}  \overline{M}^{*} ({}^3 S_1)$ and $M^{*}  \overline{M}^{*} ({}^5 S_2)$ systems}

\begin{figure}[htbp] 
	\centering
	\includegraphics[width=0.5\textwidth]{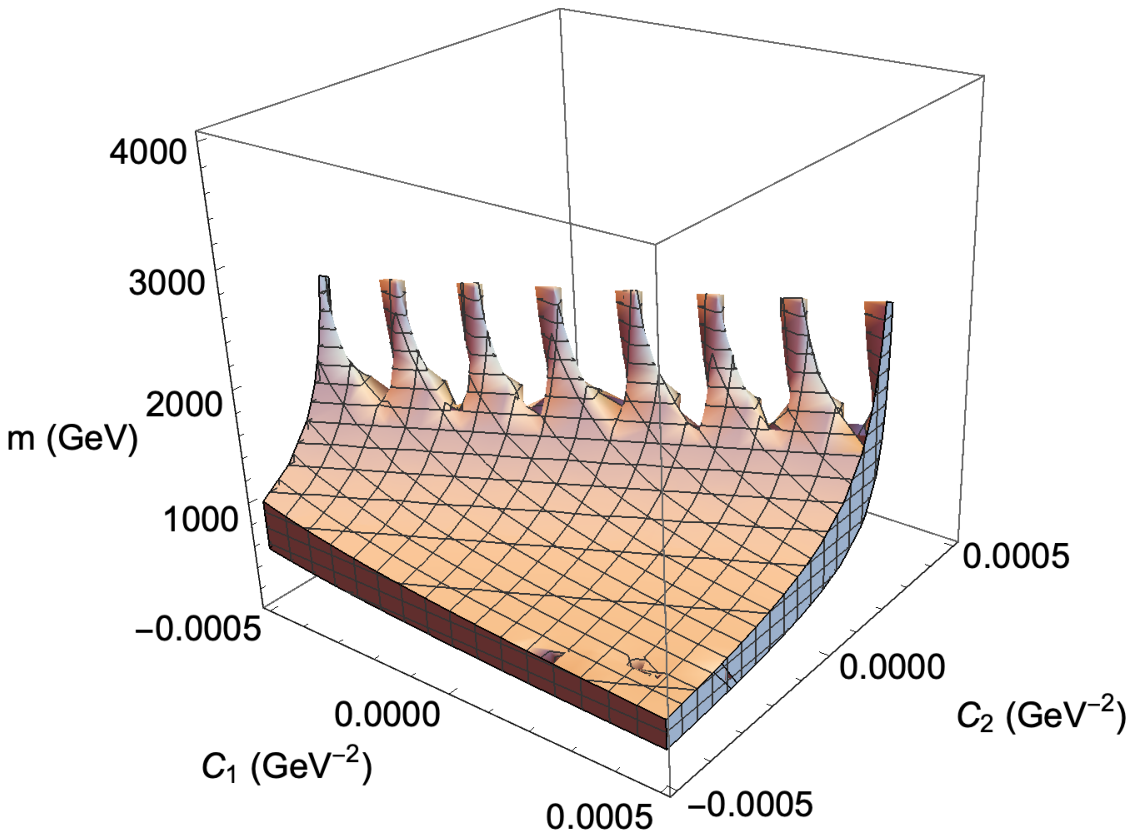}\\
	\includegraphics[width=0.5\textwidth]{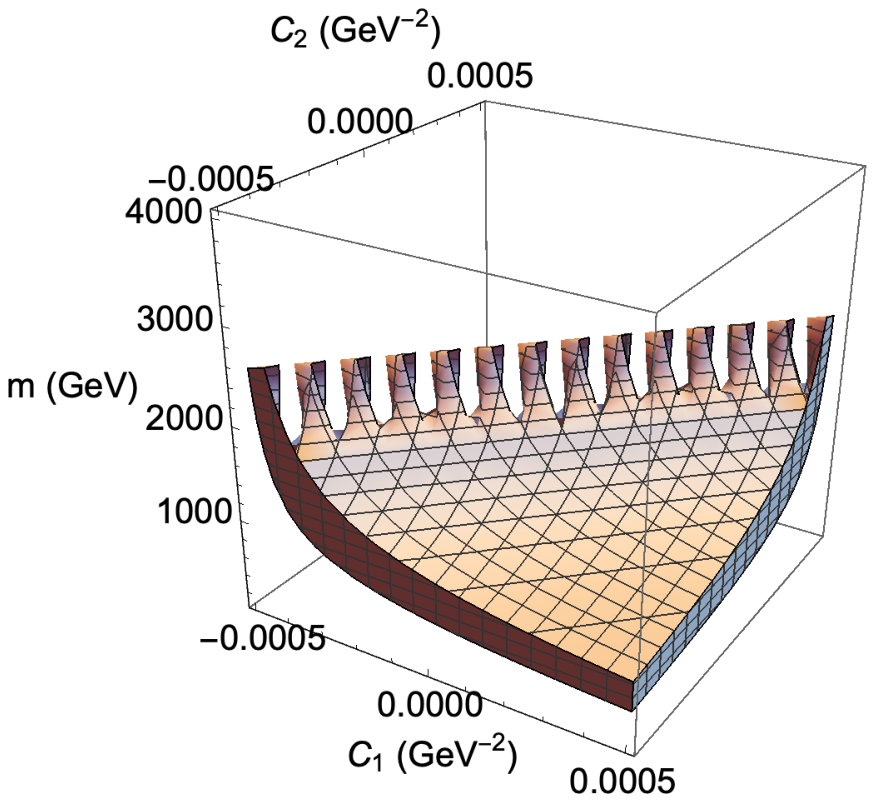}\\	\includegraphics[width=0.5\textwidth]{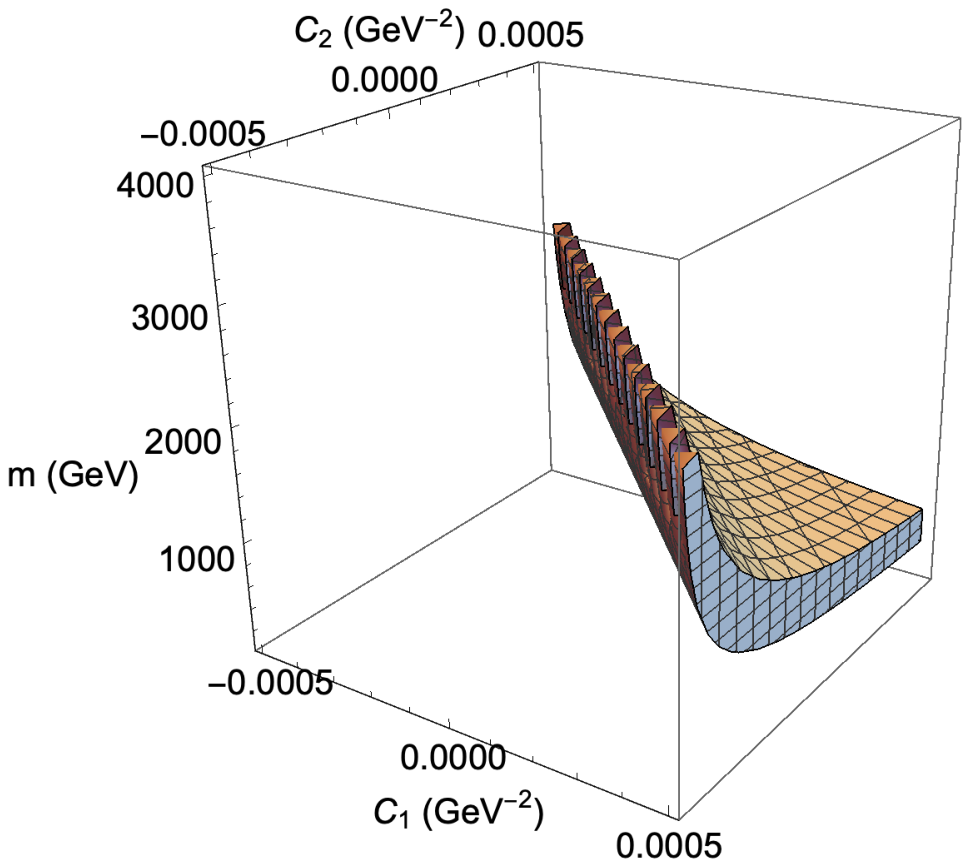}
\caption{Dependence of $(C_1, C_2)$-parameter space with the monopole mass $m_{M^{*}}$ for the $ M  \overline{M} ({}^1 S_0)$ system; shaded area represents the region in which the parameters $C_1$ and $C_2$ acquire values that allow shallow bound states. Top, middle and bottom panels represent $M^{*}  \overline{M}^{*} ({}^1 S_0), M^{*}  \overline{M}^{*} ({}^3 S_1)$ and $M^{*}  \overline{M}^{*} ({}^5 S_2)$ systems, respectively. }
	\label{FIG-spin2}
\end{figure}

For theses systems, the parameter space is richer since the transition amplitudes depend on the constants $C_1$ and $C_2$, according to corresponding channel of effective potential shown in Table~\ref{table1}. Hence we analyze the dependence of monopolium solutions on the parameters  $(C_1,C_2)$. In Fig. \ref{FIG-spin2}, we display the dependence of $(C_1,C_2)$-parameter space on the monopole's mass $m_{M^{*}}$ for the $M^{*}  \overline{M}^{*} ({}^1 S_0)$, $M^{*}  \overline{M}^{*} ({}^3 S_1)$ and $M^{*}  \overline{M}^{*} ({}^5 S_2)$ systems. Shaded areas represent the regions in which the parameters acquire values that allow bound states, with binding energy $ 0.01 m_{M^{*}} \leq E_b \leq 0.1 m_{M^{*}}$.  These plots manifest the different $C_1,C_2$-dependence of each system to get monopolium solutions.  The different cases $M^{*}  \overline{M}^{*} ({}^1 S_0)$, $M^{*}  \overline{M}^{*} ({}^3 S_1)$ and $M^{*}  \overline{M}^{*} ({}^5 S_2)$ have bound state solutions for $C_2 < C_1/2$, $C_2 < C_1 $ and $C_1+ C_2 > 0$, respectively.

\begin{figure}[htbp] 
	\centering
\includegraphics[width=0.5\textwidth]{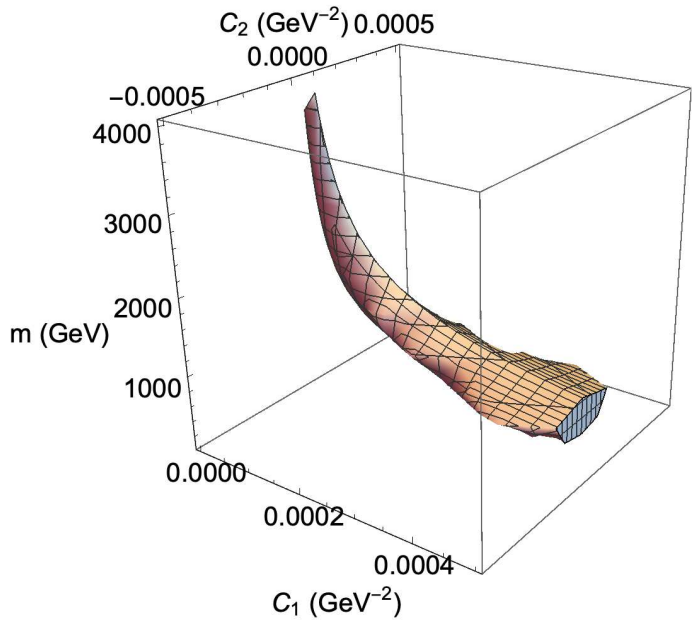}\caption{Dependence of $(C_1, C_2)$-parameter space with the monopole mass of $M^{*}  \overline{M}^{*} $ systems; shaded area represents the region of intersection in $(C_1,C_2)$-parameter space that admits shallow bound state solution for both  $M^{*}  \overline{M}^{*} ({}^1 S_0)$, $M^{*}  \overline{M}^{*} ({}^3 S_1)$ and $M^{*}  \overline{M}^{*} ({}^5 S_2)$ systems. }
	\label{FIG-spin2-2}
\end{figure}

For completeness, we also plot in Fig.~\ref{FIG-spin2-2} the region of intersection in $(C_1,C_2)$-parameter space that admits shallow bound state solution for both  $M^{*}  \overline{M}^{*} ({}^1 S_0), M^{*}  \overline{M}^{*} ({}^3 S_1)$ and $M^{*}  \overline{M}^{*} ({}^5 S_2)$ systems. This result provides values of the monopole mass and the region in parameter space that admit monopolium (shallow bound) states for all these three cases. 

\subsection{$ X_{\pm} \,\, ({}^1 S_ 0)$ systems}

We analyze monopolium solutions for these systems based on their dependence on the couplings $C_1$ and $C_2$, given in Eqs.~(\ref{LS2-XPlus}) and~(\ref{LS2-XPlus}).
Concerning the dependence on the monopole's masses, this situation violates the heavy-meson spin symmetry due to the finite heavy field masses $m_M$ and $m_{M^{*}}$ of different values. Because of the lack of experimental evidence, we assume the following relation between spin-zero and spin-one monopole masses: $m_{M^{*}} = 0.9 m_M$. 

\begin{figure}[htbp] 
	\centering
\includegraphics[width=0.5\textwidth]{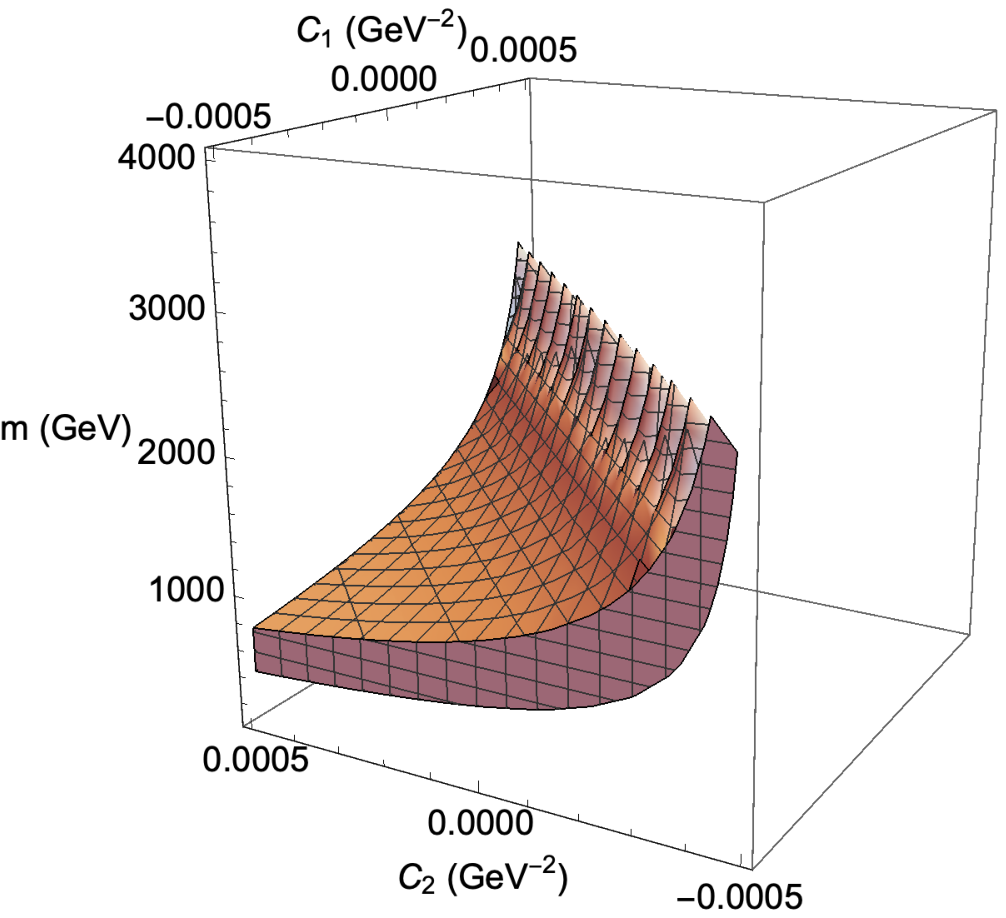}\\
\includegraphics[width=0.5\textwidth]{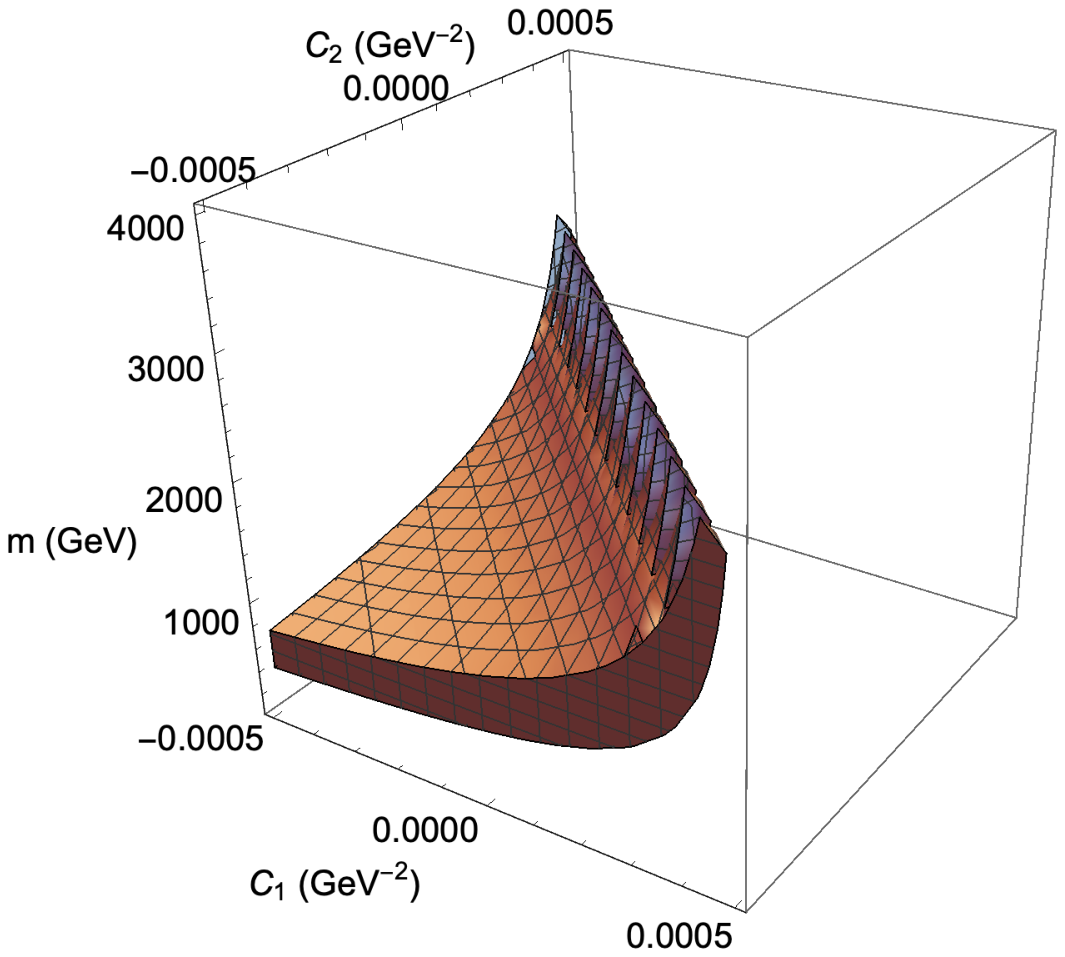}
\caption{Dependence of $(C_1, C_2)$-parameter space with the monopole mass $m_{M}$ for the $ X_{\pm} \,\, [{}^1 S_ 0]$ systems; shaded area represents the region in which the parameters $C_1$ and $C_2$ acquire values that allow shallow bound states. Top and bottom panels represent $ X_{+} \,\, ({}^1 S_ 0)$ and $ X_{-} \,\, ({}^1 S_ 0)$ systems, respectively. }
	\label{FIG-spin1}
\end{figure}

Thus, in Fig.~\ref{FIG-spin1}, we plot the dependence of $(C_1,C_2)$-parameter space on the monopole's mass $m_{M}$ for the $ X_{+} \,\, ({}^1 S_ 0)$ and $ X_{-} \,\, ({}^1 S_ 0)$ systems. Shaded areas represent the regions 
in which the parameters acquire values that allow bound states, with binding energy $E_b = 0.01 m_{M} - 0.1 m_{M}$.

These plots manifest clearly the different $C_1,C_2$-dependence of each system to obtain monopolium solutions.  The region $C_1+ C_2 > 0 $ excludes possible bound states with negative magnetic--charge conjugation. Written in another way, there are regions in $(C_1,C_2)$-parameter space that exclude shallow bound state solutions for both  $X_{+} \,\, ({}^1 S_ 0)$ and $ X_{-} \,\, ({}^1 S_ 0)$ systems.

%
   
\subsection{Estimation of Interactions}
\label{Estimation}

Although there is no experimental information of monopoles, we complete the present study by evaluating the parameter space in the light  of available analyses in the literature.  
Accordingly, in conformity with Ref.~\cite{Barrie:2016wxf}, we can use the  information for the spin-zero and spin-one monopolia of mass 5 TeV  to fix the couplings. It is worth mentioning a fundamental distinction between our approach and the one discussed in Ref.~\cite{Barrie:2016wxf}: while we consider monopolia as bound states of spin-zero and spin-one monopoles, the authors of that reference performed an evaluation of monopolium constituted of spin-1/2 monopoles. Notwithstanding, we stress that we use the values of monopolium and monopole masses to gain some insight on this issue.

We used a mass of 5 TeV for spin-zero and spin-one monopolia,  $m_M = 2624$ GeV, and $| E_b | = 248$ MeV, available in Table 1 of Ref.~\cite{Barrie:2016wxf}, as inputs of the systems  $ M  \overline{M} ({}^1 S_0)$ and $M^{*}  \overline{M}^{*} ({}^3 S_1)$ to fix the couplings. Thus, from $ M  \overline{M} ({}^1 S_0)$ state we get $C_1 = - 2.3 \times 10^{-5}$ GeV$^{-2}$. Using this value for $C_1$ in the case $M^{*}  \overline{M}^{*} ({}^3 S_1)$, we obtain $C_2 = - 4.7 \times 10^{-5}$ GeV$^{-2}$. In Table~\ref{table2} is shown the derived properties of the monopolium of mass of $5$ TeV. In this situation, there is no monopolium state with $C_{G_M} = + 1$. 

\begin{table}[ht]
  \caption{Derived properties of the monopolium of mass of $5$ TeV, and monopole mass $m_M = 2624$ GeV. We use the values of spin-zero and spin-one monopolia masses available in Table 1 of Ref.~\cite{Barrie:2016wxf} as inputs of the systems  $ M  \overline{M} ({}^1 S_0)$ and $M^{*}  \overline{M}^{*} ({}^3 S_1)$ to fix the couplings, which assume the values: $C_1 = - 2.3 \times 10^{-5}$ GeV$^{-2}$ and $C_2 = - 4.7 \times 10^{-5}$ GeV$^{-2}$. }
\begin{center}
\begin{tabular}{|c|c|c|c|c|c|c|}  \hline
State & $ M  \overline{M} ({}^1 S_0)$ & $M^{*}  \overline{M}^{*} ({}^1 S_0)$ &  $M^{*}  \overline{M}^{*} ({}^3 S_1)$  &  $M^{*}  \overline{M}^{*} ({}^5 S_2)$ & $ X_{+} \,\, ({}^1 S_ 0)$ & $ X_{-} \,\, ({}^1 S_ 0)$ \\  \hline 
$ |E_b| $ (GeV)       & 248 & 27.6 & 248 & 27.6 & - & 32.4 \\
$ a_s$ ($10^{-4}$ fm) & 2.5 &  7.4 & 2.5 & 7.4  & -   & 4.7 \\
 \hline
\end{tabular}
\end{center}
\label{table2}
\end{table}

Finally, let us dedicate ourselves to a final remark on the feasibility of detecting the monopolia suggested before. The observability depends very strongly on the binding energy and other observables like the decay width. In principle, monopolia with large binding energies  would increase the possibilities of experimental detection. This landscape has been explored in some works cited in Introduction via the choice of distinct interaction potentials and approaches~\cite{Epele,Vento,Reis,Vento:2019auh}.  However, we underline that the monopolium phenomenology is highly influenced by the dynamics assumptions. Furthermore, the absence of experimental confirmation makes the scenario also open to the alternative of relatively small bound monopolium, which is the context of this work. At this stage of experimental knowledge, given that our predictions have higher monopole masses,  the environments to observe these kinds of monopolia could be future 100 TeV colliders and astrophysical observations, through explosion or burst of $\gamma$ rays from them,  similarly to the circumstances reported in Ref.~\cite{Barrie:2016wxf}.

\section{Concluding Remarks}
\lb{Conclusions}

In summary, we have investigated the monopoles in a different theoretical perspective with respect to existing literature that allowed incorporate additional features into their interpretation and searches. We have described their basic dynamics via an effective formalism in which spin-zero and spin-one monopole interactions are taken into account through contact interactions.
In particular, the conditions of existence of bound monopole-antimonopole states have been explored from the parameter space of the coupling constants and monopole masses. This model has allowed to perform estimations of their masses, binding energies and scattering lengths as functions of interaction strength and monopole masses in a specific renormalization scheme, in which the relation between the bare coupling constants is conserved after renormalization. The possible channels as well as eigenstates of magnetic--charge conjugation have been studied and related. We have restricted ourselves to the region of relevance of contact-range interaction, in which the one-particle exchange contribution is not relevant.

The investigation of monopolium solutions for the $ M  \overline{M} ({}^1 S_0)$ system has been simpler and more direct, since it depends only on one coupling constant, which must have a negative value to admit bound state solution. For $M^{*}  \overline{M}^{*} ({}^1 S_0), M^{*}  \overline{M}^{*} ({}^3 S_1)$ and $M^{*}  \overline{M}^{*} ({}^5 S_2)$ systems, the parameter space is richer since the transition amplitudes depend on the two coupling constants $C_1,C_2$, according to corresponding channel. The results have shown a strong $C_1,C_2$-dependence of each system to get monopolium solutions, but depending on the monopole mass and the region in parameter space all these three cases admit monopolium state solutions. In the case of $ X_{\pm} \,\, ({}^1 S_ 0)$ systems, we have seen that there are regions in $(C_1,C_2)$-parameter space that exclude shallow bound state solutions for both  $X_{+} \,\, ({}^1 S_ 0)$ and $ X_{-} \,\, ({}^1 S_ 0)$ systems. 
Also, our findings have suggested that there is a specific choice of region in $(C_1,C_2)$-parameter space which admits bound state solution simultaneously for all systems, except for  monopolium state with $C_{G_M} = + 1$.

It should be highlighted that the monopole physics is highly influenced by the dynamics, which is not yet accessible. This has been stimulating the development of different ways of modelling its phenomenology. In that way, the approach reported above gives additional and unexplored aspects on the monopolonium properties by using a simplified effective model with a small number of parameters. The calculated results suggest a scenario with relatively small bound monopolia, which might be detected in future 100 TeV colliders and astrophysical observations.

Extensions and improvements can be deployed in future works. It is possible, for
example, to study the decays of these predicted molecular states into other particles. Also, the inclusion of one-particle exchange potential in order to extend the range of 
applicability of this approach is a natural development to be done.

\begin{acknowledgements}

L.M. Abreu would like to thank the Brazilian funding agencies CNPq (contracts 308088/2017-4 and 400546/2016-7) and FAPESB (contract INT0007/2016).  M de Montigny thanks the Natural Sciences and Engineering Research Council of Canada (NSERC) for financial support (grant number RGPIN-2016-04309).  P.P.A. Ouimet and M de Montigny are also grateful for funding from NSERC (grant number sAPPJ-2019-00040)

\end{acknowledgements}


%
%
%

\end{document}